\documentclass[conference]{IEEEtran}
\IEEEoverridecommandlockouts

\usepackage{cite}
\usepackage{amsmath,amssymb,amsfonts}
\usepackage{algorithmic}
\usepackage{graphicx}
\usepackage{textcomp}
\usepackage{xcolor}
\def\BibTeX{{\rm B\kern-.05em{\sc i\kern-.025em b}\kern-.08em
    T\kern-.1667em\lower.7ex\hbox{E}\kern-.125emX}}

\usepackage{romannum}

\usepackage[most]{tcolorbox}

\usepackage{tabularray}
\usepackage{booktabs}
\UseTblrLibrary{booktabs}
\SetTblrStyle{firstfoot,middlefoot,note-text}{\footnotesize\itshape}  
\usepackage{multirow} 
\usepackage[para,online,flushleft]{threeparttable}
\usepackage{array} 

\DeclareMathOperator{\randleftarrow}{\xleftarrow{\$}}
\DeclareMathOperator{\equornot}{\stackrel{?}{=}}
\DeclareMathOperator{\Sign}{Sign}
\DeclareMathOperator{\Verify}{Veri}
\DeclareMathOperator{\E}{Encry}
\DeclareMathOperator{\D}{Decry}
\DeclareMathOperator{\Hash}{H}
\DeclareMathOperator{\Linkage}{LinkGen}
\DeclareMathOperator{\Randkey}{RandKey}
\DeclareMathOperator{\Profgen}{ProfGen}   
\DeclareMathOperator{\Profver}{ProfVeri}    
\DeclareMathOperator{\Sansig}{Sant}

\usepackage{caption,graphicx,newfloat}
\DeclareCaptionType{Function}

\usepackage{subcaption}

\usepackage{amsthm}
\newtheorem{theorem}{Theorem}
\newtheorem{lemma}{Lemma}
\newtheorem{define}{Definition}
\newtheorem{game}{Security Game}

\usepackage{lmodern} 

\usepackage{url}

\begin{document}

\title{Flexible Non-interactive Short-term Implicit Certificate Generation for VANETs\\
}

\author{\IEEEauthorblockN{Rui Liu}
\IEEEauthorblockA{
\textit{Department of Computer Science} \\ \textit{University of Victoria}\\
Victoria, Canada \\
liuuvic@uvic.ca}
\and
\IEEEauthorblockN{Yun Lu}
\IEEEauthorblockA{
\textit{Department of Computer Science} \\ \textit{University of Victoria}\\
Victoria, Canada \\
yunlu@uvic.ca}
\and
\IEEEauthorblockN{Jianping Pan}
\IEEEauthorblockA{
\textit{Department of Computer Science} \\ \textit{University of Victoria}\\
Victoria, Canada \\
pan@uvic.ca}
}

\maketitle

\begin{abstract}
	A leading industry standard for secure and trusted communication in vehicular ad-hoc networks (VANETs) is the Security Credential Management System (SCMS).
	It uses anonymous certificates, functioning as pseudonyms, to preserve the privacy of vehicles.
	With the rapid development of advanced applications in VANETs, such as crowdsensing and federated learning, vehicles need to communicate with each other or infrastructures more frequently, leading to a higher demand for pseudonyms.
	However, the current approach of certificate provisioning in SCMS is not able to fully support pseudonyms, due to storage limitation, cost of connectivity establishment, and communication overhead of certificate downloading. To tackle this challenge, we propose a non-interactive approach for SCMS, allowing vehicles themselves to generate short-term key pairs and anonymous implicit certificates. Our evaluation and comparison with previous work show that our solution not only effectively reduces the communication cost, but also grants vehicles greater flexibility in certificate generation and use. On the technical side, to the best of our knowledge, this is the first work which (1) applies {\em sanitizable signature} for non-interactive anonymous certificate generation, and (2) is specifically designed for SCMS, which opens up possibilities for extensions and applications in industry.
\end{abstract}


\section{Introduction}
\label{sec_1_intro}
Secure and trusted communication in vehicular ad-hoc networks (VANETs) is a topic of recent research interest~\cite{brecht2018security,barreto2020schnorr, ercan2022enhanced, tangade2020trust, twardokus2022vehicle}. 
One leading public key infrastructure (PKI)-based solution is the Security Credential Management System (SCMS)~\cite{brecht2018security}, which has been standardized by IEEE~\cite{IEEE1609}. In SCMS, certificate authorities (CAs) generate and issue anonymous certificates to vehicles, which are used to verify the public keys of vehicles and provide message integrity and authenticity. The process is called \textit{certificate provisioning}. 
Vehicles usually obtain a batch of certificates simultaneously and change pseudonyms (i.e., certificates and public keys) on demand.

The rapid development of VANET techniques promotes advanced applications~\cite{gao2022vehicle,da2022cloud,liu2020airq,liu2023crs,xi2023rsus}. In crowdsensing, vehicles equipped with sensors can collect and upload real-time environmental data for specific tasks, such as traffic management~\cite{gao2022vehicle}, map updating~\cite{da2022cloud} and air quality monitoring~\cite{liu2020airq}. In federated learning, vehicles perform as workers to train models locally and exchange parameters with neighbor vehicles or road-side units (RSUs)~\cite{liu2023crs}. 
Safety applications also have a promising future. For example, vehicles can collect real-time traffic information and broadcast emergency messages to avoid road accidents~\cite{xi2023rsus}. 
A noticeable situation is that, to facilitate these applications, the amount of vehicle-to-everything (V2X) communication is drastically increased. 
Considering that vehicles are recommended to change pseudonyms after a certain number of messages (e.g., every 100 messages~\cite{etsi2018intelligent}), these applications, in turn, urgently increase the demand for more certificates.

Two common solutions in SCMS to address this challenge are the following: (1) downloading more certificates from the CA (through RSUs or cellular networks) each time; (2) downloading certificates more frequently. However, these solutions may be problematic~\cite{brecht2018security,ercan2022enhanced}: a) storing a very large number of certificates may not be feasible due to the limited memory storage of vehicles; b) frequently establishing connectivity and downloading certificates is expensive considering the cost of RSU deployment and cellular network access; c) 
the high mobility of vehicles may lead to unreliable communication and considerable delays.

In addition, there is a lack of flexibility in the current design of SCMS. The system manager only defines the same certificate provisioning model (e.g., at least 20 certificates are valid simultaneously within one week~\cite{brecht2018security}) for all the registered vehicles. 
Considering that a) vehicles may have different driving time and patterns in real life; b) vehicles may have different privacy concerns; and c) communication requirements vary with different applications, an identical model is not sufficient and can lead to waste of certificates or lack of pseudonyms.
Thus, a practical design should allow each vehicle to have its own personalized provisioning model.


To achieve the above desiderata, we propose a flexible \textbf{no}n-\textbf{in}teractive \textbf{s}hort-term certificate generation (NOINS) approach for SCMS. 
Following the recommendation in SCMS and a recent work, SIMPL~\cite{barreto2020schnorr}, implicit certificates are adopted rather than the traditional explicit certificates, which is more communication-efficient.
With this approach, vehicles can generate short-term implicit certificates from each CA-issued certificate, without interaction with the CA or RSUs. 
It not only reduces the number of communication rounds but also provides more flexibility in generating and using certificates. 
Our main contributions are as follows.
\begin{itemize}
	\item We propose a new approach, {\em NOINS}, and non-trivially apply sanitizable signature, for generating flexible non-interactive short-term certificates. A vehicle can generate short-term pseudonyms by itself, which avoids frequently establishing connectivity or downloading certificates from CAs or RSUs. Compared with SCMS, SIMPL, and a recent work~\cite{akil2023non}, we greatly reduce communication cost.
	\item Under our proposed approach, vehicles can personalize their certificate generation, e.g., change pseudonyms according to their driving habits and privacy requirements. 
	This flexibility circumvents the issue of limited storage and prevents both the waste and the lack of certificates. 
	\item The generated short-term public keys and certificates are unlinkable, providing vehicle privacy and message authentication. In addition, NOINS provides immutability, fraud-resistance and unforgeability.
	\item 
    NOINS is designed on top of the well-accepted SCMS infrastructure, and can be directly integrated with it. There are many interesting problems that can be further explored in the new paradigm. We give an overview of these future directions in Section~\ref{sec_6_dis}.
\end{itemize}

In the rest of the paper, in Section~\ref{sec_2_pre}, we introduce the related techniques and works. The system model and the threat model are given in Section~\ref{sec_2_model}. In Section~\ref{sec_3_design}, we present NOINS in details. In Section~\ref{sec_4_eva}, we evaluate our work with theoretical analysis and simulations. Our evaluation results show that our proposed approach achieves a clear reduction in communication cost for vehicles over previous work~\cite{brecht2018security, barreto2020schnorr,akil2023non}. In Section~\ref{sec_5_sec}, we analyze, prove and compare the security achievements. A discussion of interesting research problems in this new paradigm is given in Section~\ref{sec_6_dis}, followed by a conclusion in Section~\ref{sec_7_con}.

\section{Related Work}
\label{sec_2_pre}
\subsection{SCMS}
SCMS, as a popular vehicular public key infrastructure (VPKI), received much attention these years.
Typically, it adopts the elliptic curve Qu-Vanstone (ECQV) implicit certification model. This model involves scalar multiplication of EC points in public key verification but does not support pre-computing. 
Barreto~{\em et al.}~\cite{barreto2020schnorr} propose an alternative for SCMS. The proposed Schnorr-based implicit certification (SIMPL) approach enables pre-computing and improves the efficiency, especially on the vehicles' side. The certificate provisioning process in SCMS with SIMPL is briefly introduced in Section~\ref{subsec:sys} and given in Table~\ref{tab:approachs}. 

\subsection{Sanitizable Signature}
One important underlying technique of NOINS is the sanitizable signature, which is first proposed by Ateniese~{\em et al.}~\cite{ateniese2005sanitizable}. It allows an authorized party, called sanitizer, to modify the message signed by the signer but keeps the signature still valid. A typical way to achieve it is adopting chameleon hash functions~\cite{ateniese2005sanitizable}. However, due to the same hash values, the sanitized messages of the same document can be linked, which is a crucial shortcoming of this method.

Brzuska~{\em et al.}~\cite{brzuska2010unlinkability} first introduce the notion of unlinkability in sanitizable signatures. To achieve unlinkability, the authors adopt group signatures as the underlying signature scheme so that the sanitized messages have different signatures. Unfortunately, we cannot adopt it because a) having a large group of vehicles share the same group key leads to insecurity in certificate generation; and b) defining each vehicle and the CA as a group leads to the linkability of vehicles.
Another line of works is based on re-randomized keys. Fleischhacker~{\em et al.}~\cite{fleischhacker2016efficient} use a perfectly re-randomized secret key to sign the editable part of the message so that the sanitized messages, with different public keys, are unlinkable. 
However, the unchanged public key of the sanitizer should be revealed to prove the authenticity of the re-randomized public keys. Thus, it only guarantees that a sanitized message cannot be linked to an original message, but the messages belonging to the same sanitizer are still linkable. 

In addition, sanitizable signature schemes are first proposed for database or document management, so that particular properties are usually taken into consideration~\cite{bultel2019efficient,bossuat2021unlinkable}.
These properties, including invisibility, transparency and accountability, are not necessary for NOINS design but will increase the burden of the scheme. 
In summary, existing works on sanitizable signatures cannot be directly adopted for short-term certificate generation in VANETs.

\subsection{Self-changing Pseudonyms}
There have been some efforts in self-changing the pseudonyms of vehicles. Qi~{\em et al.}~\cite{qi2022pseudonym} propose a pseudonym-based certificateless authentication scheme. In this scheme, RSUs generate a list of partial keys for vehicles and periodically publish a related value, with which vehicles can generate key pairs on their side. 
Yang~{\em et al.}~\cite{yang2023scalable} propose a self-agent pseudonym management scheme where vehicles generate short-term pseudonyms and send them to the server for activation.
However, different from NOINS, these works are not truly non-interactive. Akil~{\em et al.}~\cite{akil2023non} propose a truly non-interactive scheme based on non-interactive zero-knowledge proofs and Camenisch-Lysyanskaya (CL) signatures. 
It allows vehicles to broadcast messages with a self-updated pseudonym and a CA-issued attribute-based certificate. However, a) it does not support message encryption and signing; b) it has high communication and computation costs; and c) it cannot be integrated with SCMS directly. We further evaluate and compare our work with the most related works~\cite{brecht2018security, barreto2020schnorr,akil2023non} in Section~\ref{sec_5_sec}.

To the best of our knowledge, this is the first work applying (and further revising) sanitizable signatures in the non-interactive anonymous certificate generation problem and specifically designed for SCMS.

\section{System Model and Threat Model}
\label{sec_2_model}
\subsection{System Model}
\label{subsec:sys}
\begin{figure}[tb!]
	\centering
	\includegraphics[width=1.0\linewidth]{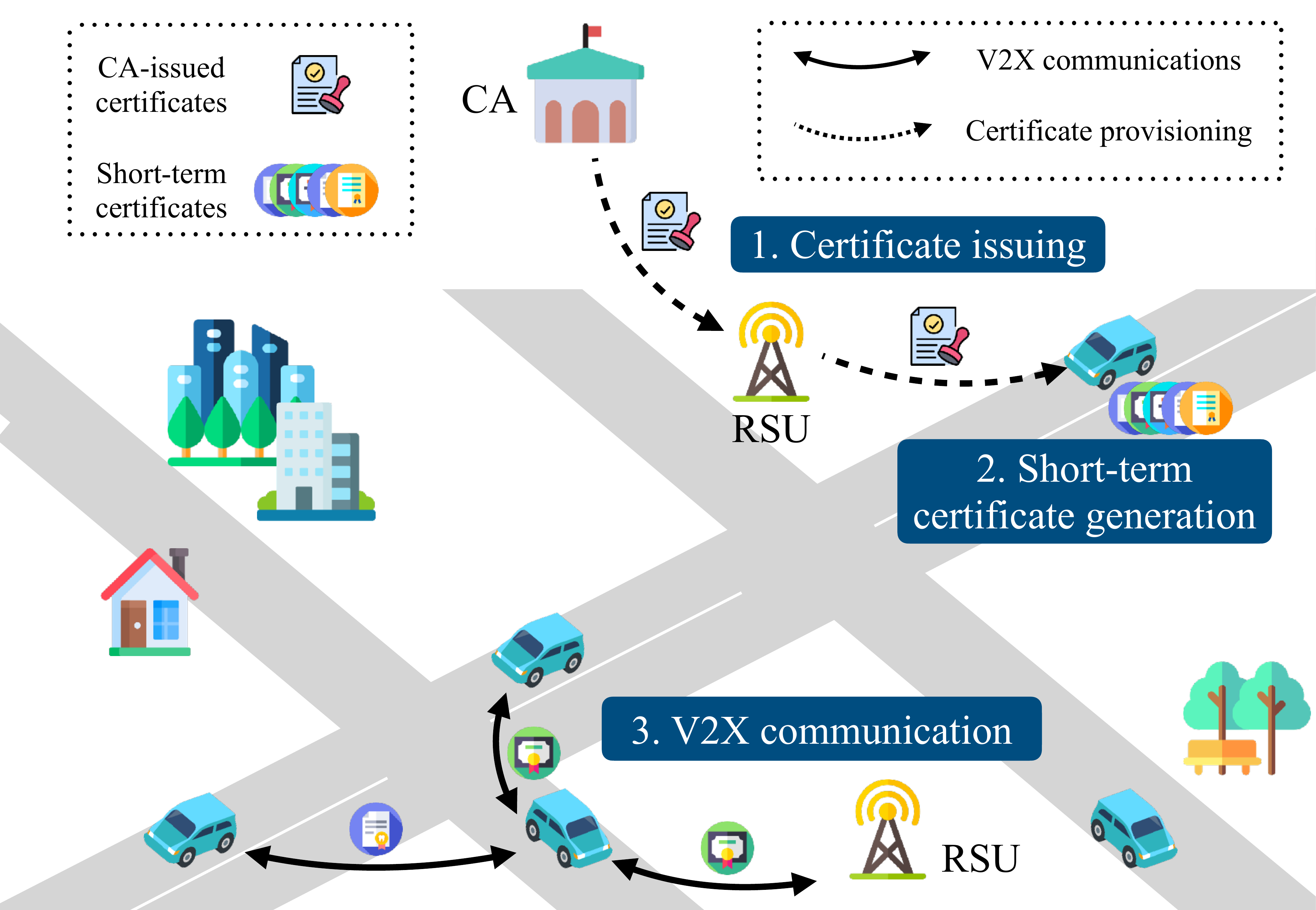}
	\caption{System model} 
	\vspace{-3mm}
	\label{fig:model}
\end{figure}

As shown in Figure~\ref{fig:model}, we formally define three entities:
\begin{itemize}
	\item \textit{Vehicles}: vehicles participating in concrete tasks (such as data collection in crowdsensing and model training in federated learning) have more V2X communications and high demand for anonymous certificates. 
	They request implicit certificates $\mathit{cert}$ from a certificate authority and generate short-term ones on demand. The initial key pair generated by a vehicle itself is $(x, X)$, called caterpillar keys. The associated signature key pair of $\mathit{cert}$, used for V2X communication, is denoted as $(\mathit{skv}, \mathit{pkv})$.
	\item \textit{CA}: the certificate authority is responsible for issuing implicit certificates $\mathit{cert}$ to vehicles. The key pair of CA is denoted as $(\mathit{skc}, \mathit{pkc})$. Note that, there are different certificate authorities in SCMS. These authorities work together to ensure that no individual component knows the entire set of data that can be used to track a vehicle. For simplicity, in this paper, we only use one CA as a logical component to represent the authorities in SCMS.
	\item \textit{RSUs}: RSUs are wireless communication devices or base stations on the roadside. All the messages transmitted between vehicles and the CA are relayed by RSUs.
\end{itemize}

SCMS employs the unified butterfly key expansion process for certificate provisioning. To be specific, a vehicle first generates a caterpillar key pair $(x\xleftarrow{\$} \mathbb{Z}_q, X\leftarrow x\cdot g)$ where $\xleftarrow{\$}$ indicates randomly sample a value from a group. $g$ is the generator of an elliptic curve group $\mathbb{G}$ of prime order $q$. Upon receiving certification requests from vehicles, a registration authority (RA) first expands every $X$ to multiple $\hat{X_i}$ (called cocoon keys) with a function $f$ and then shuffles all $\hat{X_i}$ from all vehicles. CA is responsible for issuing a certificate, with which the associated public key can be reconstructed, for each $\hat{X_i}$. This process allows each vehicle to request a batch of certificates (and key pairs) while avoiding CA from linking the vehicle with the issued certificate batch. 
The proposed solution, NOINS, works with the unified butterfly key expansion process but focuses on the processes afterwards, so that we omit the description of the related contents in the following sections. We refer readers to~\cite{brecht2018security, barreto2020schnorr} for more details.
For simplicity, we leave out the subscript $i$ and use $\hat{X}$ to denote a cocoon key of $X$. Note that all operations in group $\mathbb{G}$ and group $\mathbb{Z}_q$ are made $(\mod q)$ in this paper.

\subsection{Threat Model}
\label{subsec:threat}
We assume each party in the system is either honest or corrupt, where a party is honest if it both follows the protocol and can be trusted with secret information. A corrupt party may be ``semi-honest". It follows the protocol but might be interested in, e.g., inferring the private key of other parties, pretending to be some legitimate vehicle, forging a certificate, deanonymizing other vehicles, or inferring their trajectories. 
We assume the CA is honest. 
We assume corrupt RSUs and registered vehicles (called internal adversaries) are semi-honest.  
There may exist external adversaries, who have the ability to capture and inspect the messages transmitted in the system and certificates of other legitimate vehicles.
The major security properties we consider are described as follows:
\begin{itemize}
	\item \textit{Immutability}: when generating short-term certificates from the CA-issued certificate, vehicles cannot modify metadata such as certificate expiration dates.
	\item \textit{Anonymity}: certificates cannot be used to reveal vehicles' sensitive information or real identities.
	\item \textit{Unlinkability}: short-term certificates and associated public keys should be unlinkable, i.e., corrupt parties cannot link two certificates (or two public keys) to the same vehicle and thus infer its trajectory.
	\item \textit{Fraud-resistance}: an adversary can not pose as another vehicle, even with access to this vehicle's (legitimate) certificates.
	\item \textit{Unforgeability}: an adversary cannot forge a new valid certificate, even with access to valid certificates from legitimate vehicles.
\end{itemize}

Considering the applications in VANETs mentioned in Sections~\ref{sec_1_intro} and~\ref{subsec:sys}, we assume the vehicles who participate in these tasks would like to expose their real identities to CA to obtain task rewards. We further discuss this assumption in Section~\ref{sec_6_dis}. 
Physical attacks, such as tracing a vehicle by its color and speed, are out of scope.

\section{NOINS Design}
\label{sec_3_design}

The flexible non-interactive short-term implicit certificate generation approach is shown in Table~\ref{tab:approachs}. For each cocoon public key $\hat{X}$, CA generates a reconstruction value $\mathit{rcv}$ and an implicit certificate $\mathit{cert}$. With $\mathit{rcv}$, a signature public key, $\mathit{pkv}$ can be reconstructed for the vehicle. To distribute the CA-issued certificate, a message $m_\mathrm{I2V}$ is sent to the vehicle. The message is encrypted with the vehicle's cocoon public key $\hat{X}$ so that only the vehicle can decrypt it. After obtaining and verifying the CA-issued certificate, the vehicle generates short-term certificates $\mathit{cert}_j$ with the sanitization technique. A new short-term key pair $(\mathit{skv}_j, \mathit{pkv}_j)$ can be generated with each $\mathit{cert}_j$. For the sake of security and privacy, a short-term sanitization key pair $(\mathit{sks}_j, \mathit{pks}_j)$ is used in this process. 
In a V2X communication, the receiver can verify the authenticity of $ \mathit{pks}_j$ and then reconstruct $\mathit{pkv}_j$ from $\mathit{cert}_j$. The verification of $\mathit{pkv}_j$ will be implicitly conducted when it is used, i.e., when verifying a signature signed with the corresponding $\mathit{skv}_j$. 
Both the short-term sanitization public keys and the short-term signature public keys are unlinkable. 

Our scheme allows for flexible, on-demand short-term certificate generation, which can e.g., be personalized to each vehicle based on their driving habits and privacy requirements. For example, a vehicle only driving on weekends typically need fewer certificates than a vehicle driving every day. A vehicle can pre-generate a batch of short-term ones and then generate new ones when suffering a shortage of pseudonyms. A vehicle uploading real-time sensory data needs more frequent pseudonym changing than a vehicle that uploads data less frequently. 

Frequently used notations are summarized in Table~\ref{tab:note}. The processes of certificate generation and provisioning in SCMS with traditional anonymous explicit certificates (explicit approach for short) and with implicit SIMPL are also given in Table~\ref{tab:approachs} for easy comparison. 
Now we give a detailed description of the proposed NOINS approach.


\begin{table}[t!]
	\centering
	\caption{Frequently used notations}
	\label{tab:note}
	\begin{threeparttable}
	\begin{tblr}{colspec={X[1,l]X[3,l]}, rowspec={Q[m]Q[m]}}
		\toprule
		Notation~\tnote{$\ast$} & Description \\
		\midrule 
		$(x,X)$  & The caterpillar key pair of a vehicle\\
		$(\hat{x},\hat{X})$ &  A cocoon key pair of a vehicle\\
		$(\mathit{skv},\mathit{pkv})$ & The signature key pair of a vehicle\\
		$(\mathit{skv}_j,\mathit{pkv}_j)$ & A short-term signature key pair of a vehicle\\
		$(\mathit{skc,\mathit{pkc}})$ & The key pair of CA \\
		$cert$ & A CA-issued certificate associated with $\mathit{pkv}$\\
		$cert_j$ & A short-term certificate associated with $\mathit{pkv}_j$\\
		$(\mathit{sks},\mathit{pks})$ & The sanitization key pair issued by CA\\
		$(\mathit{sks}_j,\mathit{pks}_j)$ & A short-term sanitization key pair\\
		$\mathit{rcv}$ & The public key reconstruction value\\
		$h^1,h^2,h^2_j$ & Hash values \\
		$sig, sig^1,sig^2$& Digital signatures \\
		$sig^2_j$& The sanitized $sig^2$ with $\mathit{sks}_j$\\
		$\mathit{meta}$ & The metadata of a digital certificate\\
		$\mathit{lv}$ & The linkage value of a vehicle \\
		$\mathit{slv}_j$ & A short-term linkage value of a vehicle \\
		\bottomrule 
	\end{tblr}
 \begin{flushleft} 
	\begin{tablenotes}
		\footnotesize
		\item[$\ast$] \textit{In this paper, all superscripts are notations (to distinguish the parameters of the same type) rather than exponents. All operations in group $\mathbb{G}$ and group $\mathbb{Z}_q$ are expressed additionally. In all key pairs, the first entry is private key while the second one is public key.}
	\end{tablenotes}	    \end{flushleft}
	\end{threeparttable}
\end{table}

\begin{table*}[ht!]
	\centering
	\caption{Certificate generation and provisioning process}
	\label{tab:approachs}
	\begin{threeparttable}
		\begin{tblr}{colspec={Q[c,1.2cm]|Q[c,4.5cm]|Q[c,0.7cm]|Q[c,5.6cm]|Q[c,0.7cm]|X[c]}, rowspec={Q[m]|Q[m]|Q[m]|Q[m]|Q[m]|Q[m]}}

			\hline 
			\rule{0pt}{2.5ex} & CA & $\rightarrow$~\tnote{$\dagger$} & Vehicle &$\rightarrow$~\tnote{$\ddagger$} & Receiver\\
			\hline
			
			\hfil SCMS \newline (explicit) \newline\cite{brecht2018security, barreto2020schnorr}&
			\hfil $r \randleftarrow \mathbb{Z}_q$ \newline\null\hfil 
			$\mathit{pkv} \leftarrow \hat{X}+r\cdot g$\newline\null\hfil 
			$\mathit{sig}\leftarrow \Sign_\mathit{skc}(\{\mathit{pkv},\mathit{meta}, \mathit{lv}\})$\newline\null\hfil 
			$\mathit{cert}\leftarrow\{\mathit{pkv}, \mathit{meta},\mathit{sig}\}$ \newline
			$m_\mathrm{I2V} \leftarrow \E_{\hat{X}}(\{\mathit{cert},\mathit{r}\})$
			
			& \SetCell[r=3]{c}$m_\mathrm{I2V} $& \hfil$\hat{x}\leftarrow f(x) $\newline\null\hfil 
			$\{\mathit{cert},r\}\leftarrow \D_{\hat{x}}(m_\mathrm{I2V})$\newline\null\hfil 
			$\Verify_\mathit{pkc}(\mathit{cert})$ \newline\null\hfil 
			$\mathit{skv}\leftarrow \hat{x}+\mathit{r}$
			\newline
			$\mathit{pkv} \equornot \mathit{skv}\cdot g $
			& \hfil$\mathit{cert}$\newline $\mathit{pkv}$ & $\Verify_\mathit{pkc}(\mathit{cert})$
			\\
			\cline{1-2}\cline{4-6}
			
			\hfil SCMS \newline (implicit) \newline (SIMPL)\newline\cite{barreto2020schnorr}&
			\hfil $r \randleftarrow \mathbb{Z}_q$ \newline\null\hfil 
			$\mathit{rcv} \leftarrow \hat{X}+r\cdot g$\newline\null\hfil 
			$\mathit{cert}\leftarrow\{\mathit{rcv}, \mathit{meta},\mathit{lv}\}$ \newline\null\hfil 
			$h\leftarrow \Hash(\mathit{cert}, \mathit{pkc})$ \newline\null\hfil 
			$\mathit{sig} \leftarrow r+h\cdot \mathit{skc}$ \newline
			$m_\mathrm{I2V} \leftarrow \E_{\hat{X}}(\{\mathit{cert},\mathit{sig}\})$
			
			& & \hfil$\hat{x}\leftarrow f(x) $\newline\null\hfil 
			$\{\mathit{cert},\mathit{sig}\}\leftarrow \D_{\hat{x}}(m_\mathrm{I2V})$\newline\null\hfil 
			$h\leftarrow \Hash(\mathit{cert}, \mathit{pkc})$ \newline\null\hfil 
			$\mathit{skv}\leftarrow \hat{x}+\mathit{sig}$
			\newline
			$\mathit{pkv}\leftarrow\mathit{skv}\cdot g \equornot \mathit{rcv}+h\cdot \mathit{pkc}$
			& $\mathit{cert}$ & \hfil$h\leftarrow \Hash(\mathit{cert}, \mathit{pkc})$\newline
			$\mathit{pkv}\leftarrow \mathit{rcv}+h\cdot  \mathit{pkc}$
			\\
			\cline{1-2}\cline{4-6}
			
			NOINS& \hfil $r^1, r^2\randleftarrow \mathbb{Z}_q$\newline\null\hfil 
			$\mathit{rcv} \leftarrow \hat{X}+r^1\cdot g+r^2\cdot g$\newline\null\hfil 
			$\mathit{cert}\leftarrow\{\mathit{rcv}, \mathit{meta},\mathit{lv}\}$ \newline\null\hfil 
			$h^1\leftarrow \Hash(\mathit{meta}, \mathit{pkc})$ \newline\null\hfil 
			$\mathit{sig}^1 \leftarrow r^1+h^1\cdot \mathit{skc}$ \newline\null\hfil 
			$h^2\leftarrow \Hash(\mathit{rcv}, \mathit{lv}, \mathit{pks})$ \newline\null\hfil 
			$\mathit{sig}^2 \leftarrow r^2+h^2\cdot \mathit{sks}$ \newline
			$m_\mathrm{I2V} \leftarrow \E_{\hat{X}}(\{\mathit{cert},\mathit{sig}^1,\mathit{sig}^2, \mathit{sks}, r^2\})$
			& & \hfil$\hat{x}\leftarrow f(x) $\newline
	    	$\{\mathit{cert},\mathit{sig}^1,\mathit{sig}^2, \mathit{sks}, r^2\} \leftarrow \D_{\hat{x}}(m_\mathrm{I2V})$\newline
	    	$(\hat{x}+\mathit{sig}^1+\mathit{sig}^2)\cdot g \equornot \mathit{rcv}+h^1\cdot \mathit{pkc}+h^2\cdot \mathit{pks}$ 
	    	\begin{tcolorbox}[colback=gray!20,enhanced,sharp corners,frame hidden,halign=center,top=0.2em,bottom=0.05em, toprule=-0.3em, bottomrule=-0.3em] To generate each $\mathit{cert}_j$:
	    	\end{tcolorbox}
	    	\hfil$\mathit{slv}_j \leftarrow\Linkage(\mathit{lv})$\newline\null\hfil 
	    	$r^3_j, r^4_j, \rho_j \randleftarrow \mathbb{Z}_q$\newline\null\hfil 
	    	$\mathit{rcv}_j \leftarrow \mathit{rcv}+r^3\cdot g$\newline\null\hfil 
	    	$(\mathit{sks}_j,\mathit{pks}_j)\leftarrow\Randkey(\mathit{sks},\mathit{pks},\rho_j)$\newline\null\hfil 
	    	$\{\mathit{com}_j, \mathit{resp}_j \}\leftarrow \Profgen(r^4_j, \rho_j)$\newline\null\hfil 
	    	$\mathit{sig}^2_j \leftarrow \Sansig(\mathit{rcv}_j, \mathit{slv}_j, \mathit{sks}_j,\mathit{pks}_j, r^2)$\newline\null\hfil 
	    	$\mathit{cert}_j\leftarrow\{\mathit{rcv}_j, \mathit{meta},\mathit{slv}_j\}$ \newline\null\hfil 
			$\mathit{skv}_j\leftarrow \hat{x}+\mathit{sig}^1+\mathit{sig}^2+r^3_j$
			\newline
			$\mathit{pkv}_j\leftarrow\mathit{skv}_j\cdot g $ & $\mathit{cert}_j$\newline $\mathit{pks}_j$ \newline $\mathit{com}_j$\newline$\mathit{resp}_j $
			& $\Profver(\mathit{com}_j, \mathit{resp}_j, \linebreak \mathit{pks}_j, \mathit{pks})$\newline
			$h^1\leftarrow \Hash(\mathit{meta}, \mathit{pkc})$\newline
			$h^2_j\leftarrow \Hash(\mathit{rcv}_j, \mathit{slv}_j, \mathit{pks}_j)$\newline
			$\mathit{pkv}_j\leftarrow \mathit{rcv}_j+h^1\cdot \mathit{pkc}+h^2_j\cdot \mathit{pks}_j$
			\\ 
			\hline
		\end{tblr}
	 \begin{flushleft} 
		\begin{tablenotes}
			\footnotesize
			\item[$\dagger$] \textit{The message sent from CA to a vehicle through an RSU. }
			\item[$\ddagger$] \textit{Values sent in V2X communications for authentication.}
		\end{tablenotes}
	\end{flushleft} 
	\end{threeparttable}
\end{table*}

\subsection{CA-issued Certificate Generation}
\label{subsec:ca_iss}
For each cocoon public key $\hat{X}$, CA is responsible for generating a certificate $\mathit{cert}$, which is valid in a certain period of time.

As shown in Table~\ref{tab:approachs}, CA first picks two random values $r^1$ and $r^2$ and calculates the reconstruction value $\mathit{rcv}$ with $\hat{X}$. Besides $\mathit{rcv}$, $\mathit{cert}$ contains some metadata $meta$ (for example, the certificate format, the responsible authority, and the expiration time) and a linkage value $\mathit{lv}$ as well. The system-defined metadata should not include any user-identifiable information. $\mathit{lv}$ is a novel concept in SCMS and used for efficient revocation of certificate. It is unique in each CA-issued certificate. We refer readers to~\cite{brecht2018security} for more details. Considering the applications of VANETs, it can also be used to distribute task rewards to the corresponding vehicles. 

Two hash values are generated with a hash function $\Hash$. 
The private key of CA, $\mathit{skc}$, is used to generate $\mathit{sig}^1$ with the first hash $h^1$ so that no one can sanitize $\mathit{sig}^1$ without $\mathit{skc}$. It is considered as the immutable part. 

The second hash $h^2$ is used in generating $\mathit{sig}^2$ with a sanitization private key $\mathit{sks}$. The sanitization key pair is generated by CA in advance as $(\mathit{sks}\in \mathbb{Z}_q, \mathit{pks}\leftarrow \mathit{sks}\cdot g)$. It is shared with the vehicle so that the vehicle can sanitize the editable part, i.e., $\mathit{sig}^2$, for generating new certificates. A key security problem is that, if each vehicle has a unique sanitization key pair shared with CA, it should not be exposed to other vehicles, otherwise unlinkability is lost. In Section~\ref{subsec:st_gen}, we will show that this restriction cannot be achieved. To circumvent this, in NOINS, we consider the CA-issued sanitization key pair to be identical to a large number of vehicles, for example, to all the vehicles registered in a city or all vehicles whose license plates share the same prefix. The principle is that revealing the sanitization public key would not reveal any sensitive information about a vehicle. An expiration date can be set for each sanitization key pair as well.

When issuing $\mathit{cert}$ to the vehicle, CA encrypts the message ($\{\mathit{cert}, \mathit{sig}^1, \mathit{sig}^2,\mathit{sks}, r^2\}$) with $\hat{X}$. It guarantees that even though the sanitization key is shared among vehicles, only the vehicle with the corresponding private key $\hat{x}$ can obtain $r^2$ and generate short-term certificates from $\mathit{cert}$. $\E$ is an asymmetric encryption function such as the elliptic curve integrated encryption scheme (ECIES) while $\D$ is the corresponding decryption function.

\subsection{Short-term Certificate Generation}
\label{subsec:st_gen}
The vehicle first gets the cocoon private key $\hat{x}$ associated with $\hat{X}$, as mentioned in Section~\ref{subsec:sys}. After decrypting the message $m_\mathrm{I2V}$, the vehicle can verify the CA-issued certificate $\mathit{cert}$. The correctness can be proved straightforwardly, as shown in \eqref{equ:certi}.
\begin{equation}
\label{equ:certi}
	\begin{split}
	& (\hat{x}+\mathit{sig}^1+\mathit{sig}^2)\cdot g \\
	& = \hat{X}+ (r^1+h^1\cdot \mathit{skc})\cdot g +(r^2+h^2\cdot \mathit{sks})\cdot g \\
	& =(\hat{X}+ r^1\cdot g + r^2 \cdot g ) + h^1\cdot \mathit{skc}\cdot g+ h^2\cdot \mathit{sks}\cdot g\\
	& =\mathit{rcv}+h^1\cdot \mathit{pkc} +h^2\cdot \mathit{pks}.
	\end{split}
\end{equation}

The vehicle can generate the short-term certificates from each $\mathit{cert}$ on demand, which provides much flexibility.

Note that the linkage value $\mathit{lv}$ is linkable if it is reused in different short-term $\mathit{cert}_j$ where $j\in \{1,2,\cdots, n_\mathrm{cs}\}$. $n_\mathrm{cs}$ is a system pre-defined upper limitation. Thus, the vehicle generates a short-term linkage value $\mathit{slv}_j$ from $\mathit{lv}$ for each $\mathit{cert}_j$. To achieve that, we adopt the same approach with SCMS (specifically, the last step of generating $\mathit{lv}$ from pre-linkage values)~\cite{brecht2018security} in Function~\ref{func:link}. 
In short, it is a pseudorandom function (such as AES) in the Davies-Meyer mode. $\mathrm{AES}_{\mathit{lv}}(a)$ denotes encrypting $a$ with taking $\mathit{lv}$ as the key. $\mathit{ID}_\mathrm{CA}$ is the identity string associated with CA. $[a]_{t_\mathrm{sv}}$ denotes the $t_\mathrm{sv}$ significant bytes of bit-string $a$. For the sake of security, $\mathit{lv}$, as the secret key, should not be revealed to other entities. 

\begin{Function}[ht]
	\centering
	\fbox{
	\parbox{\dimexpr\linewidth-2\fboxsep-2\fboxrule\relax \small}{\centering
		$\mathit{slv}_j \leftarrow   [\mathrm{AES}_{\mathit{lv}}(\mathit{ID}_\mathrm{CA}\|j) \oplus (\mathit{ID}_\mathrm{CA}\|j)]_{t_\mathrm{sv}}$ \\
		Output $\mathit{slv}_j $}}
	\caption{$\Linkage(\mathit{lv})$\label{func:link}}
\end{Function}

The vehicle re-randomizes $\mathit{rcv}$ with a random number $r^3$ to obtain a new reconstruction value $\mathit{rcv}_j$. 
We expect to use a sanitization key pair $(\mathit{sks},\mathit{pks})$ to generate a new signature $\mathit{sig}^2_j$ for $\mathit{rcv}_j$ and $\mathit{slv}_j$. The process is called \textit{sanitizing}. 
Now we answer the previous question raised in Section~\ref{subsec:ca_iss}, i.e., why we require CA to issue the same $(\mathit{sks},\mathit{pks})$ to a great many vehicles.
In a V2X communication, to reconstruct the short-term signature public key of a message sender, the message receiver must get $\mathit{pks}$. If $\mathit{pks}$ is unique to the sender, the unlinkability is broken. Adversaries can link different $\mathit{cert}_j$ with the same $\mathit{pks}$.

There is another problem that must be considered: allowing different vehicles to use the same private sanitization key in certificate generation is not secure. To tackle the problem while upholding unlinkability, we do not directly use the CA-issued $(\mathit{sks},\mathit{pks})$ in sanitizing. A unique sanitization key pair $(\mathit{sks}_j,\mathit{pks}_j)$ will instead be generated from it and used in each $\mathit{cert}_j$ generation. With this design, each vehicle has its own short-term private sanitization keys and revoking $\mathit{pks}_j$ does not break unlinkability.
To this end, similar to the work in~\cite{fleischhacker2016efficient}, the vehicle re-randomizes the shared sanitization key pair $(\mathit{sks},\mathit{pks})$ by Function~\ref{func:randk}~\footnote{$\mathit{pks}_j$ can also be calculated as $\mathit{sks}\cdot g$ but calculating $\mathit{pks}+\rho_j \cdot g$ is more efficient for the whole process. The result of $\rho_j \cdot g$ can be stored and used in the upcoming Function~\ref{func:zero} so that one point multiplication operation on $\mathbb{G}$ is saved.}. 

\begin{Function}[ht]
	\centering
	\fbox{
		\parbox{\dimexpr\linewidth-2\fboxsep-2\fboxrule\relax \small}{\centering
			$\mathit{sks}_j \leftarrow \mathit{sks}+\rho_j$ \\
			$\mathit{pks}_j \leftarrow \mathit{pks}+\rho_j \cdot g$ \\
			Output $(\mathit{sks}_j , \mathit{pks}_j )$}}
	\caption{$\Randkey(\mathit{sks},\mathit{pks},\rho_j)$\label{func:randk}}
\end{Function}

Intuitively, the process of sharing $(\mathit{sks},\mathit{pks})$ can be considered as CA distributes the power of short-term certificate generation to legitimate, registered vehicles.
Then, the remaining task for the vehicle who uses $(\mathit{sks}_j,\mathit{pks}_j)$ instead is to prove that they indeed get the power from CA. This can be achieved by a zero-knowledge proof of the relationship between $(\mathit{sks}_j,\mathit{pks}_j)$ and $(\mathit{sks},\mathit{pks})$. To be specific, the vehicle computes a commitment $\mathit{com}_j$ and a response $\mathit{resp}_j$ by Function~\ref{func:zero}, which is a non-interactive version of the standard sigma (challenge-response) protocol~\cite{fiat1986prove}. Unlike~\cite{fleischhacker2016efficient}, there is no need to expose the caterpillar public key $X$ or cocoon public key $\hat{X}$ of the vehicle, which avoids the violation of unlinkability. 

\begin{Function}[ht]
	\centering
	\fbox{
		\parbox{\dimexpr\linewidth-2\fboxsep-2\fboxrule\relax \small}{\centering
			$\mathit{com}_j\leftarrow r^4_j \cdot g$ \\
			$\mathit{cha}_j\leftarrow \Hash(g, \mathit{com}_j, \rho_j\cdot g)$ \\
			$\mathit{resp}_j\leftarrow r^4_j + \mathit{cha}_j\cdot \rho_j$\\
			Output $\mathit{com}_j$ and $\mathit{resp}_j$}}
	\caption{$\Profgen(r^4_j, \rho_j)$\label{func:zero}}
\end{Function}

After setting up these keys, $\mathit{sig}^2$ is sanitized to $\mathit{sig}^2_j$ by Function~\ref{func:sani}. The main idea is that the vehicle can re-calculate a new $\mathit{sig}^2$ (i.e., $\mathit{sig}^2_j$) for a short-term $\mathit{cert}_j$ with different $\mathit{rcv}$, $\mathit{slv}$ and $\mathit{pks}$ (i.e., $\mathit{rcv}_j$, $\mathit{slv}_j$ and $\mathit{pks}_j$). Thus, a short-term key pair $(\mathit{skv}_j,\mathit{pkv}_j)$ can be generated and associated with $\mathit{cert}_j$. The security is guaranteed by the secrecy of $r^2$, $r^3$ and $\mathit{sks}_j$. The short-term certificate $\mathit{cert}_j$ contains the immutable part, $\mathit{meta}$, and the sanitized part, $\mathit{rcv}_j$ and $\mathit{slv}_j$. It is still an implicit certificate so that $\mathit{pkv}_j$ can be reconstructed from it and verified implicitly.

\begin{Function}[ht]
	\centering
	\fbox{
		\parbox{\dimexpr\linewidth-2\fboxsep-2\fboxrule\relax \small}{\centering
			$h^2_j\leftarrow \Hash(\mathit{rcv}_j, \mathit{slv}_j, \mathit{pks}_j)$ \\
			$\mathit{sig}^2_j \leftarrow r^2+h^2_j\cdot \mathit{sks}_j$ \\
			Output $\mathit{sig}^2_j$}}
	\caption{$\Sansig(\mathit{rcv}_j, \mathit{slv}_j, \mathit{sks}_j,\mathit{pks}_j, r^2)$\label{func:sani}}
\end{Function}

\subsection{Short-term Certificate Using in V2X Communications}
In a V2X communication, a vehicle, as a sender, signs its message $m_\mathrm{V2X}$ with its short-term private key $\mathit{sks}_j$. It needs to provide its certificate $\mathit{cert}_j$, $\mathit{pks}_j$, $\mathit{com}_j$ and $\mathit{resp}_j$ to the message receiver for authentication. 

The receiver first checks the expiration date of the certificate and the validity of $\mathit{pks}$.
It then verifies the authenticity of $\mathit{pks}_j$ by Function~\ref{func:very}. If it outputs Succeed, the receiver generates $h^1$ and $h^2_j$ and reconstructs the short-term public key of the sender, i.e., $\mathit{pkv}_j$. 
The signed message $m_\mathrm{V2X}$ can be verified with $\mathit{pkv}_j$. Meanwhile, $\mathit{pkv}_j$ is implicitly verified because a successful verification of $m_\mathrm{V2X}$ indicates that the sender indeed holds $\mathit{sks}_j$.

\begin{Function}[ht]
	\centering
	\fbox{
		\parbox{\dimexpr\linewidth-2\fboxsep-2\fboxrule\relax \small}{\centering
			$\mathit{cha}_j\leftarrow \Hash(g, \mathit{com}_j, \mathit{pks}_j-\mathit{pks})$ \\
			If $\mathit{resp}_j \cdot g = (\mathit{pks}_j-\mathit{pks})\cdot \mathit{cha}_j + \mathit{com}_j$ \\
			Output Succeed \\
			Otherwise, output Fail}}
	\caption{$\Profver(\mathit{com}_j, \mathit{resp}_j, \mathit{pks}_j, \mathit{pks})$\label{func:very}}
\end{Function}

\section{Performance Evaluation}
\label{sec_4_eva}
In this section, following the evaluation paradigm of SIMPL~\cite{barreto2020schnorr}, we compare the performance of NOINS, SIMPL and a recent non-interactive approach~\cite{akil2023non} (Akil's approach for short). 
SIMPL works better than the original implicit certificate-based approach of SCMS~\cite{barreto2020schnorr} so that we do not compare NOINS with the original implicit SCMS again.
Considering that explicit certificates are still widely used in real life, we also involve the traditional explicit approach of SCMS for a clear comparison.
We set $\mathbb{G}$ with the elliptic curve Secp256k1. A 256-bit prime order $q$ is adopted to provide 128-bit security. We choose SHA-256 as the general one-way hash function. RSA-2048 with SHA-256, which is supported in the widely known X.509 certificates, is adopted as the digital signature scheme in the explicit approach.
As recommended by SCMS, ECIES with a 256-bit curve is selected as the encryption function in $m_\mathrm{I2V}$. The parameters used in Akil's approach are based on the recommendation of the idemix protocol~\cite{IBMprotocol}.

We start by measuring the computation time. The experimental platform is composed of an Intel Q9550 CPU with 2.83GHz frequency, and 8GB RAM. 
Following the work in~\cite{barreto2020schnorr}, multiplications by $\mathit{pkc}$ are optimized by the Comb method (with a window width $w=8$). 
Suppose that in both the explicit approach and SIMPL, $n_\mathrm{c}$ certificates are provided for a vehicle in total. 
Suppose there are $n_\mathrm{ci}$ 
CA-issued certificates in NOINS and Akil's approach. $n_\mathrm{cs}$ short-term certificates (or pseudonyms in Akil's approach) are generated from each of them and $n_\mathrm{c}=n_\mathrm{ci}\cdot n_\mathrm{cs}$. In other words, we consider the number of total certificates (or pseudonyms) a vehicle can use to be the same for comparison, no matter whether it is a CA-issued one or a self-generated short-term one.

\begin{table}[ht!]
	\centering
	\caption{Computation cost for $n_c$ certificates (ms)}
	\label{tab:comp_1}
        \begin{threeparttable}
	\begin{tblr}{colspec={X[1,c]X[1,c]X[1,c]X[1,c]}, rowspec={Q[m]Q[m]Q[m]Q[m]Q[m]}}
		\toprule
		& CA & Vehicle & Reciever \\
		\midrule 
		Explicit~\cite{brecht2018security} & 307.76$n_\mathrm{c}$ & 11.84$n_\mathrm{c}$ & 1.98$n_\mathrm{c}$ \\
		SIMPL~\cite{barreto2020schnorr} & 28.81$n_\mathrm{c}$ & 11.85$n_\mathrm{c}$&1.99$n_\mathrm{c}$ \\
            Akil's approach~\cite{akil2023non}~\tnote{$\S$} & 3073.49$n_\mathrm{ci}$ &  3352.07$n_\mathrm{ci}$ + 11553.26$n_\mathrm{c}$ & 1297.10$n_\mathrm{c}$ \\
            NOINS & 37.37$n_\mathrm{ci}$&20.40$n_\mathrm{ci}$+34.10$n_\mathrm{c}$& 27.61$n_\mathrm{c}$\\
		\bottomrule 
	\end{tblr}
        \begin{flushleft} 
	\begin{tablenotes}
		\footnotesize
		\item[$\S$] \textit{We consider 10 attributes maintained in each certificate as an example. Because the approach is not comprehensively described in details by the authors, we only take the major processes into evaluation.}
        \end{tablenotes}	    \end{flushleft}
        \end{threeparttable}
\end{table}

The computation cost is given in Table~\ref{tab:comp_1}~\footnote{This is the evaluation method we use: we comprehensively analyze all the operations on different cryptography groups in these approaches. We execute each operation 1000 times with the Multiprecision Integer and Rational Arithmetic Cryptographic Library (MIRACL)~\cite{MIRACL} for average results and then calculate the total computation time. For example, to generate one certificate, CA has to conduct one point addition (0.0573ms) and one point multiplication (8.4791ms) on $\mathbb{G}$, one RSA signing (278.9789ms), and one ECIES encryption (20.2469ms), which are 307.76ms in total. Please note that all computation time can be further reduced with advanced platforms or mature toolkits in practice.}. It can be observed that NOINS has the lowest computation burden on the side of CA (recall that $n_\mathrm{ci}<n_\mathrm{c}$).
The explicit approach and SIMPL do not support self-generation of pseudonyms so that has lower computation burden on vehicles' side. NOINS and Akil's approach both hold this promising feature but NOINS is much more lightweight.
Considering the increased computing capacity and the challenges of stable communication in VANETs, it is worth obtaining more improvements on communication burden with some sacrifice of computation efficiency. It echoes our design goal, i.e., reducing the communication cost and providing more flexibility for vehicles.

To measure the communication cost of vehicles, we conduct simulations with the Network Simulator (ns)-3.34~\cite{ns3}. 
In all simulations, we model the CA-issued certificates as being sent to vehicles via RSUs. For completeness, the process of using certificates (i.e., sending a certificate and related parameters to a receiver in V2X communications) is also measured.
We consider the scenarios of a) a small city (e.g., Victoria), and b) a large city (e.g., Shanghai). It reflects different CA-to-RSU distances: around 5km and 60km, respectively. The propagation speed of the wired communication between CA and RSUs equals the speed of light, i.e., 299792.46km/s. The standard of IEEE 802.11p is adopted for V2X communication.
The distance between a vehicle to the nearest RSU to be in the range of 0m to 300m. 
According to the two-second rule, we consider the distance between two vehicles is in the range of 10m to 100m in communication. The Transmission Control Protocol (TCP) is adopted for all communications.

We consider 16-byte metadata as an example.
$\mathit{slv}$ is set as 9 bytes. In the explicit approach and SIMPL, 9-byte $\mathit{lv}$ is used as suggested by SCMS~\cite{brecht2018security}. Differently, in NOINS, we set $\mathit{lv}$ as 16 bytes to provide 128-bit security in AES. 
we consider $n_\mathrm{cs}=50$ and $n_\mathrm{c}=\{500, 1000, 3000\}$ as representatives.
We consider each time a batch of 20 CA-issued certificates is sent together to the vehicle in all simulations according to the CAR 2 CAR Communication Consortium (C2C-CC) model~\cite{bissmeyer2011generic}.  
For simplicity, we omit the detailed description and directly give the results. 

Table~\ref{tab:comu_1} shows the end-to-end delay required for obtaining $n_\mathrm{c}$ certificates (or pseudonyms in Akil's approach). Table~\ref{tab:comu_2} shows the total communication time for obtaining and using these certificates. It can be observed that our approach, NOINS, obviously saves the communication cost for vehicles. We further compare the security achievments in Section~\ref{sec_5_sec}.

\begin{table}[t!]
	\centering
	\caption{Communication time required for obtaining $n_\mathrm{c}$ certificates (in seconds)}
	\label{tab:comu_1}
	\begin{tblr}{colspec={X[1,c]X[1,c]X[1,c]X[1,c]X[1,c]X[1,c]X[1,c]}, rowspec={Q[m]Q[m]Q[m]Q[m]Q[m]Q[m]}}
		\toprule
		&  \SetCell[c=3]{c} Small city & & & \SetCell[c=3]{c}  Large city & & \\
		\midrule
	$n_\mathrm{c}$	& 500 & 1000 & 3000 &  500 & 1000 & 3000  \\
		\midrule 
		Explicit & 0.49  & 0.98 & 2.95 & 0.65 & 1.31 & 3.93\\
		SIMPL & 0.18 & 0.37 & 1.10 & 0.24 & 0.49 & 1.46 \\
            Akil's & 0.03 & 0.03 & 0.09 & 0.04 & 0.04 & 0.12\\
            NOINS & 0.01 & 0.01 & 0.04 & 0.02 & 0.02 & 0.05 \\
		\bottomrule 
	\end{tblr}
\end{table}

\begin{table}[t!]
	\centering
	\caption{Total communication time for obtaining and using $n_\mathrm{c}$ certificates (in seconds)}
	\label{tab:comu_2}
	\begin{tblr}{colspec={X[1,c]X[1,c]X[1,c]X[1,c]X[1,c]X[1,c]X[1,c]}, rowspec={Q[m]Q[m]Q[m]Q[m]Q[m]Q[m]}}
		\toprule
		&  \SetCell[c=3]{c} Small city & & & \SetCell[c=3]{c}  Large city & & \\
		\midrule
	$n_\mathrm{c}$	& 500 & 1000 & 3000 &  500 & 1000 & 3000  \\
		\midrule 
		Explicit & 0.83  & 1.65 & 4.96 & 0.99 & 1.98 & 5.94\\
		SIMPL & 0.33 & 0.66 & 1.99 & 0.39 & 0.78 & 2.35 \\
            Akil's & 3.59 & 7.15 & 21.44 & 3.60 & 7.15 & 21.46\\
            NOINS & 0.23 & 0.44 & 1.31 & 0.23 & 0.44 & 1.33 \\
		\bottomrule 
	\end{tblr}
\end{table}

\section{Security Analysis}
\label{sec_5_sec}
In this section, we describe and analyze the security properties NOINS provides, based on the threat model defined in Section~\ref{subsec:threat}. 

\textbf{Immutability}.
We first define the property in NOINS as follows:
\begin{define}
	The approach is immutable if there is no probabilistic polynomial-time (PPT) adversary $\mathcal{A}$, i.e., a vehicle registered in the system and trying to generate short-term certificates, can win Security Game~\ref{game:immu} with probability of success that is non-negligible in $k$ where $k$ is the security parameter.
\end{define}
\begin{game}
	\label{game:immu}
	An adversary $\mathcal{A}$, with cacoon private key $\hat{x}$, obtains $\{\mathit{cert},\mathit{sig}^1,\mathit{sig}^2, \mathit{sks}, r^2\}$ from CA where $\mathit{sig}^1=r^1+h^1\cdot \mathit{skc}$. $\mathcal{A}$ wins if it can generate $\mathit{sig}^{1}_\mathcal{A}$ with respect to a bitstring of its choice $\mathit{str}_\mathcal{A}$ and satisfying $(\hat{x}+\mathit{sig}^{1}_\mathcal{A}+\mathit{sig}^2)\cdot g = \mathit{rcv}+h^{1}_\mathcal{A}\cdot \mathit{pkc}+h^2\cdot \mathit{pks}$ where $h^{1}_\mathcal{A}=\Hash(\mathit{str}_\mathcal{A})$.
\end{game}

Immutability is guaranteed with the security of $\mathit{sig}^1$ as stated in Theorem~\ref{theorem:immu}.
\begin{theorem}
	\label{theorem:immu}
	Immutability holds in NOINS if the generation of $\mathit{sig}^1$ is forgery-resistant.
\end{theorem}
\begin{proof}
	Suppose a PPT adversary $\mathcal{A}$ wins Security Game~\ref{game:immu} with a non-negligible probability $p$.
	Then it holds that $(\hat{x}+\mathit{sig}^{1}_\mathcal{A}+\mathit{sig}^2)\cdot g = \mathit{rcv}+h^{1}_\mathcal{A}\cdot \mathit{pkc}+h^2\cdot \mathit{pks}$, which leads to $\mathit{sig}^{1}_\mathcal{A}\cdot g = r^1\cdot g + h^{1}_\mathcal{A}\cdot \mathit{pkc}$.
	It indicates that $\mathcal{A}$ is able to generate a forged signature for message $\mathit{str}_\mathcal{A}$ with the Schnorr signature scheme, with respect to public key $\mathit{pkc}$, with probability $p$.
\end{proof}

With Lemma~\ref{lemma:schnorr}~\cite{pointcheval2000security} and Theorem~\ref{theorem:immu}, the immutability holds in NOINS.

\begin{lemma}
	\label{lemma:schnorr}
	The Schnorr signature is existentially unforgeable under chosen-message attacks (EU-CMA) in the Random
	Oracle Model (ROM) under the Discrete Logarithm (DL) assumption.
\end{lemma}

\textbf{Anonymity and unlinkability}. Anonymity is guaranteed by adopting the same principle as SCMS, i.e., avoiding sensitive identity information in the metadata.
We focus on discussing the unlinkability of the shared information in V2X communication: a) $\mathit{cert}_j=\{\mathit{rcv}_j, \mathit{meta},\mathit{slv}_j\}$, b) $\mathit{pks}_j$, and c) $\mathit{com}_j$ and $\mathit{resp}_j$. 

\begin{define}
	\label{define:unlink}
	Two items of interest are unlikable if no PPT adversary $\mathcal{A}$ can win Security Game~\ref{game_unlink} with non-negligibly (in security parameter $k$) larger or smaller probability than the probability imposed by its a-priori knowledge~\footnote{A-priori knowledge~\cite{pfitzmann2010terminology} is the background knowledge adversaries have before running Security Game~\ref{game_unlink}, such as the total number of registered vehicles.}. 
\end{define}

\begin{game}
	\label{game_unlink}
	An adversary $\mathcal{A}$ obtains two items $a$ and $b$ of interest. 
	It applies a judge function $\{0,1\}\leftarrow \mathrm{Jug}(a,b)$ to determine if (\romannum{1}) $a$ and $b$ are generated from the same vehicle or with the same value, or (\romannum{2}) $a$ is generated from $b$.
	$\mathcal{A}$ wins if $\mathrm{Jug}(a,b)$ outputs the correct answer.
\end{game}

\begin{theorem}
	\label{theorem:unlink}
	Unlinkability holds in NOINS for any pair of distinct items from the set $\{\mathit{cert}_j, \mathit{pks}_j, \mathit{com}_j, \mathit{resp}_j\}$ and any pair of $\{\mathit{cert}_j, \mathit{cert}\}$ for $\forall j\in N_\mathrm{cs}$. $N_\mathrm{cs}=\{1,2,\cdots, n_\mathrm{cs}\}$.
\end{theorem}
\begin{proof}
	Suppose the adversary $\mathcal{A}$ has a-priori knowledge $K$.
	
	$\mathit{rcv}_j$ is generated with a random variable $r^3_j$, which is picked uniformly from $\mathbb{Z}_q$. It implies that for any $\mathit{rcv}$, all $\mathit{rcv}_j$ look uniformly random as well, as stated in Lemma~\ref{lemma:unlink_rcv}. 
	\begin{lemma}
		\label{lemma:unlink_rcv}
		Fix $\mathbb{Z}_q$, $g$ and $N_\mathrm{cs}$. For any $\mathit{rcv}$, we have
		\begin{equation}
		\begin{split}
		\{    \{r^3_j \randleftarrow \mathbb{Z}_q \mid j \in N_\mathrm{cs}: \{ \mathit{rcv}+r^3_j\cdot g \mid j \in N_\mathrm{cs} \}\\  \equiv
		\{    \{\mathit{rcv}_j \randleftarrow \mathbb{Z}_q \mid j \in N_\mathrm{cs}: \{\mathit{rcv}_j  \mid j \in N_\mathrm{cs} \}
		\end{split}
		\end{equation}
		where $\equiv$ denotes the identical distribution.
	\end{lemma}

Thus, except $K$, $\mathcal{A}$ cannot get any new knowledge helpful for $\mathrm{Jug}(\mathit{rcv}_j,\mathit{rcv}_{j^\prime})$ or $\mathrm{Jug}(\mathit{rcv}_j,\mathit{rcv})$ where $j$, $j^\prime$ $\in N_\mathrm{cs}$. 
Suppose $K^\prime$ is the a-posteriori knowledge of $K$, along with the two items of interest ($(\mathit{rcv}_j,\mathit{rcv}_{j^\prime})$ or $(\mathit{rcv}_j,\mathit{rcv})$), in Security Game~\ref{game_unlink}. 
We have $\lvert \mathrm{Pr}(\mathcal{A} \mathrm{~wins}\mid K)-\mathrm{Pr}(\mathcal{A} \mathrm{~wins}\mid K^\prime) \rvert \leq \mathrm{negl}(k)$ where $\mathrm{Pr}$ is the probability of an event happens. $\mathrm{negl}$ is a negligible function.

A similar argument holds for $\mathit{slv}_j$, $\mathit{pks}_j$, $\mathit{com}_j$ and $\mathit{resp}_j$: $\mathit{slv}_j$ of the same vehicle is generated as a truncated value of AES cipher with the Davies-Meyer construction~\cite{brecht2018security}. 
Similar with Lemma~\ref{lemma:unlink_rcv}, every re-randomized key pair $(\mathit{sks}_j, \mathit{pks}_j)$ with a uniformly chosen randomness has an identical distribution with $(\mathit{sks}, \mathit{pks})$~\cite{fleischhacker2016efficient}. $\mathit{com}_j$ and $\mathit{resp}_j$ are generated with uniformly and randomly picked $r^4_j$ and $\rho_j$. Thus, they all hold unlinkability. 
\end{proof}

\textbf{Fraud-resistance}. We first define fraud-resistance as follows:
\begin{define}
	\label{define:fraud}
	The proposed approach is fraud-resistant if for any PPT adversary $\mathcal{A}$, with key pair $(\mathit{skv}_\mathcal{A}, \mathit{pkv}_\mathcal{A})$, the success probability of winning Security Game~\ref{game:fraud} is negligible in $k$ (security parameter).
\end{define}

\begin{game}
	\label{game:fraud}
	The adversary $\mathcal{A}$ has access to an oracle which outputs signatures $\theta=\Sign_{\mathit{skv}_j}(m_\mathrm{V2X})$ signed by a valid vehicle, with respect to the short-term public keys $\mathit{pkv}_j$ and certificates $\mathit{cert}_j$ of that vehicle.
	$\mathcal{A}$ generates a signature $\theta_\mathcal{A}$ for a message $\mathit{str}$ of its choice. Given a function $\Verify_a(b,c)$ which outputs 0 if the the signature $b$ is not valid for the message $c$ under the public key $a$ and outputs 1 otherwise, $\mathcal{A}$ wins the game if $\Verify_{\mathit{pkv}_j}(\theta_\mathcal{A}, \mathit{str})$ outputs 1.
\end{game}

Based on Definition~\ref{define:fraud}, the security of NOINS relies on the digital signature scheme adopted for communication in VANETs. We formally give the theorem and proof as follows.
\begin{theorem}
	The proposed approach is fraud-resistant if the digital signature scheme $\mathcal{DS}$, composed of a signing function $\Sign$ and a verification function $\Verify$, used in V2X communication is forgery-resistant.
\end{theorem}
\begin{proof}
	The generated short-term certificate is used to reconstruct a public key. The public key is implicitly verified when the signature of the communicated message is verified. Thus, the certificate is only valid for the entity that has the associated private key.
	
	If a PPT adversary $\mathcal{A}$, without the associated private key, can win Security Game~\ref{game:fraud} with a non-negligible probability of success $p$, it can forge signatures in $\mathcal{DS}$. 
\end{proof}

\textbf{Unforgeability}.
We first define the property as follows.
\begin{define}
	\label{define:unfor}
	The short-term certificate satisfies unforgeability if any PPT adversary $\mathcal{A}$ (i.e., an unregistered vehicle) cannot win Security Game~\ref{game:unfor} with a non-negligible probability of success $p$ in $k$ (security parameter).
\end{define}
\begin{game}
	\label{game:unfor}
	An adversary $\mathcal{A}$ is not registered in the system but has access to an oracle $\mathcal{O}^1$ which outputs short-term certificates of legitimate vehicles, $\mathit{cert}_j=\{\mathit{rcv}_j, \mathit{meta},\mathit{slv}_j\}$. A worse case is that $\mathcal{A}$ also has access to an oracle $\mathcal{O}^2$ which outputs valid sanitization key pairs $(\mathit{sks}_\mathcal{A}, \mathit{pks}_\mathcal{A})$ and the associated $\mathit{com}_\mathcal{A}$ and $\mathit{resp}_\mathcal{A}$.
	
	$\mathcal{A}$ wins if it can generate a certificate $\mathit{cert}_\mathcal{A}$ (with respect to some reconstruction value $\mathit{rcv}_\mathcal{A}$, some meta data $\mathit{meta}_\mathcal{A}$ and some sub-linkage value $\mathit{slv}_\mathcal{A}$ of its choice) and a key pair $(\mathit{skv}_\mathcal{A}, \mathit{pkv}_\mathcal{A})$ satisfying a) $\mathit{pkv}_\mathcal{A} = \mathit{skv}_\mathcal{A}\cdot g$, and b) $\mathit{pkv}_\mathcal{A} = \mathit{rcv}_\mathcal{A}+\Hash(\mathit{meta},\mathit{pkc})\cdot \mathit{pkc}+\Hash(\mathit{rcv}_\mathcal{A}, \mathit{slv}_\mathcal{A}, \mathit{pks}_\mathcal{A})\cdot \mathit{pks}_\mathcal{A}$. 
\end{game}

\begin{theorem}
	The proposed short-term certificate satisfies unforgeability with respect to Definition~\ref{define:unfor}~\footnote{
For unforgeability, we assume adversaries do not collude (i.e., share information with each other). Collusion trivially breaks unforgeability because an adversary may generate a new certificate with the secret information, such as $\mathrm{sig}^1$ shared by another vehicle.
As for the other security properties, collusion between two adversaries cannot bring any advantages for winning the corresponding game. 
See Section~\ref{sec_6_dis} for future directions on relaxing the non-collusion assumption.}.
\end{theorem}
\begin{proof}
	An adversary $\mathcal{A}$ is able to generate $\mathit{rcv}_\mathcal{A}$ from existing $\mathit{rcv}_j$ with any $r^3_\mathcal{A}$ it chooses. However, to get a short-term private key $\mathit{skv}_\mathcal{A}$ to win the game, both $\mathit{sig}^1$ and $r^2$ are required. If $\mathcal{A}$ wins the game with probability $p$, then it can make a forgery on the underlying scheme with probability $p$. With Lemma~\ref{lemma:schnorr}, the underlying scheme is provably secure, yielding a contradiction.
\end{proof}

\textbf{Comparison}. NOINS is specifically designed for SCMS. It achieves the essential security and privacy requirements of SCMS and SIMPL. Differently, it can provide more flexibility and address the challenges of certificate provisioning. 

Both NOINS and Akil's approach support non-interactive self-generated pseudonyms. They both provide anonymity, conditional anonymity/non-repudiation (i.e., CA can deanonymize vehicles), unlinkability, fraud-resistance, unforgeability, non-repudiation, and offline verification (i.e., a message can be verified without interacting with third parties)~\cite{akil2023non}. 
Differently, with Akil's approach, a vehicle can only have one pseudonym each time period, which can prevent Sybil attacks but has less flexibility. 
NOINS, following the recommendation of SCMS~\cite{brecht2018security}, allows a vehicle to hold multiple pseudonyms simultaneously based on its demands. In addition, Akil's approach only generates pseudonyms but does not change the key pair of a vehicle. Unlinkable message encryption and PKI-based message signing are not supported, which limits its application in VANETs. Our approach, NOINS, not only supports these important functions but is also much more lightweight than Akil's approach.
\section{Discussion}
\label{sec_6_dis}
As a new paradigm in SCMS, there are many interesting problems that can be further studied.

\textbf{Limiting the power of vehicles}. The research on the underlying technique, sanitizable signature, provides many possibilities for an enhanced design of NOINS. 
One extension is to limit the power of sanitizers. For example, limiting the number of versions that they can generate on one document~\cite{canard2010extended}, i.e., allowing us to impose limitations on the number of pseudonyms a vehicle can generate with each CA-issued certificate. 
One possible method is to adopt the technique of cryptographic accumulator. 
The remaining challenge is how to avoid the exposure of linkable information in this process, which may violate the property of unlinkability.

\textbf{Malicious vehicles and collusion attacks}. Another valuable research direction is to consider fully malicious (i.e., arbitrarily deviates from the protocol) instead of semi-honest vehicles. For example, one assumption in this work is that vehicles would like to provide their real identities (i.e., the real $\mathit{slv}_j$ which can be linked to identities according to the design in~\cite{brecht2018security}) to CA for task rewards. Although it is reasonable for most real-world users, finding a solution to enforce it can provide more security, especially when we consider an adversary who does not care about rewards but only aims at taking some illegitimate actions and escaping from identity tracing. 

Moreover, while we assume non-collusion for unforgeability, an interesting question is whether this property is possible with limited collusion. We know, for example, unforgeability holds if adversaries only share public information with each other. 

\textbf{Templates for personalized models}. NOINS provides much flexibility for vehicles. Vehicles can work within various personalized models for certificate generation and use. Although there is actually no standard, a system manager can provide some templates for vehicles.
This idea is widely used in industry; for example, computer security software companies (such as CORTEX) usually provide playbooks for users in security orchestration and automation. 
With these templates, we not only allow users to personalize their own models but also make the solution more user-friendly.

We give some template examples from different perspectives: a) a vehicle can generate enough short-term certificates with the home network before traveling outside. 
It can discard each certificate after one-time use.
This is suitable for a vehicle with good storage capability and strong concerns of network security and privacy. 
b) In a crowdsensing task of air quality monitoring~\cite{liu2020airq}, 
vehicles are asked to upload sensory data to the nearest RSU every 15 minutes.
Considering the risk of exposing their trajectories in this process, vehicles should change their pseudonyms every 15 minutes as well. c) According to the recommendation of the ETSI standard~\cite{etsi2018intelligent}, vehicles can change pseudonyms every 5 minutes, 100 messages, or every 500m. It is more suitable when the communication is intensive and the privacy concern is not as strong. 
Defining different templates should take many concrete factors into consideration and has many possibilities. 

Overall, applying the idea of sanitizable signatures in non-interactive short-term implicit certificate generation is a novel paradigm for SCMS, with obvious advantages in communication efficiency and flexibility. Interesting research problems can be further explored.

\section{Conclusion}
\label{sec_7_con}
In this paper, we propose a flexible non-interactive short-term implicit certificate generation approach. After obtaining CA-issued certificates, a vehicle can generate short-term key pairs and certificates on demand. It is achieved by adopting and modifying the technique of sanitizable signature. Vehicles can determine the time of generation, the number of certificates, and the frequency of pseudonym changing. It not only avoids the waste or lack of pseudonyms but also significantly reduces the communication cost on vehicles' side. The proposed approach is designed on top of SCMS so that is naturally suitable for it. As a new paradigm, there are many extensions and applications to be explored. 

\bibliographystyle{IEEEtranS}
\bibliography{IEEEabrv,article}

\end{document}